\documentclass[english,pra,preprint,showpacs]{revtex4}
\usepackage[T1]{fontenc}
\usepackage[latin1]{inputenc}
\usepackage{babel}
\usepackage{graphics}
\usepackage{amsmath}
\usepackage{amssymb}

\input epsf

\begin{document}
\title{Competition among molecular fragmentation channels described with
Siegert channel pseudostates}

\author{Edward L. Hamilton and Chris H. Greene}

\affiliation{Department of Physics and JILA, University of Colorado,
 Boulder, Colorado 80309-0440}

\begin{abstract}
To describe multiple interacting fragmentation continua, we develop a method
in which the vibrational channel functions obey outgoing wave Siegert boundary
conditions. This paper demonstrates the utility of the Siegert approach, which
uses channel energy eigenvalues that possess a negative imaginary part. The
electron scattering energy in each such channel is rotated upward, giving it
an equal and opposite imaginary part. This permits a natural inclusion of
vibrational continua without requiring them to appear as explicit channels in
the scattering matrix. Calculations illustrate the application of this theory
to photoionization, photodissociation, and dissociative recombination.

\end{abstract}
\maketitle

In this Letter, we propose a general method for describing coupling between
electronic and dissociative continua, based on a Siegert pseudostate basis
representation of the vibrational degree of freedom. The underlying rationale
for this idea traces back to the original recognition by Kapur and Peierls
\cite{Kapur1938} that the narrow resonances of a scattering spectrum can be
described in terms of a complex energy eigenstate, with the imaginary part of
the energy defining a resonance width parameter. This proposal was further
developed by Siegert \cite{Siegert1939} in two important ways. First, Siegert
demonstrated that a set of eigenstates of any Hamiltonian could be chosen to
satisfy pure incoming or outgoing wave boundary conditions in the asymptotic
limit. Second, and of importance to our work, Siegert's derivation allowed for
overlapping resonances of arbitrary width, giving a smooth background term
in the cross-section. Siegert eigenstates formally correspond to S-matrix
poles in the complex plane. Sharply resonant features associated with bound
states can be identified with poles lying on the real axis, while broad
background scattering can be described by closely spaced eigenstates with
finite imaginary parts that serve as a discretized approximation to the true
continuum.

Traditionally, the use of Siegert states has been complicated by the
nonlinearity of the associated eigenvalue problem. Because the wavenumber
appears linearly in the boundary condition but quadratically (as the energy)
in the eigenvalue, the eigenproblem is quadratic, and in the past could only
be solved iteratively. Tolstikhin \emph{et al.}
\cite{Tolstikhin1998,Tolstikhin1997} recently demonstrated how this difficulty
may be circumvented using finite range \emph{Siegert pseudostates}.

Initially, the true asymptotic boundary condition is replaced by a finite
range approximation,
\begin{equation}
\label{eq:spsbc}
\left( \frac{d}{dR}-ik \right) \phi(R) \bigg|_{R = R_{0}} = 0,
\end{equation}
\noindent
where \(R_{0}\) is a value beyond which the value of the potential is
negligible. We seek a solution expanded in terms of some primitive basis set
\begin{equation}
\label{eq:expansion}
\phi(R)=\sum_{j=1}^{N} c_{j} \, y_{j}(R), \qquad 0 \le R \le R_{0}.
\end{equation}
\noindent
Here $N$ is the dimension of our basis, and we have selected a non-orthogonal
B-spline basis for the $y_{j}(R)$. Inserting this into the Schr\"odinger
equation premultiplying by $y_{j'}$, and employing the boundary value
\eqref{eq:spsbc}, we find a matrix equation for the coefficients $c_{j}$
\begin{equation}
\label{eq:eqforcj}
\begin{split}
& \sum_{j=1}^{N} \left(
 \frac{1}{2} \int_{0}^{R_{0}} \frac{d y_{j'}}{dR} \frac{d y_{j}}{dR} dR -
 \frac{ik}{2} y_{j'}(R_{0}) y_{j}(R_{0}) \right. \\
 & \left. \quad + \int_{0}^{R_{0}} y_{j'}(R)\mu [V(r)-E] y_{j}(R) dR \right)
 c_{j} = 0. \\
\end{split}
\end{equation}
\noindent
Note that we have used a Green's theorem identity before substituting in the
boundary condition, and that the Hamiltonian has been multiplied through by
the reduced mass \(\mu\). Written more concisely in matrix notation, we have a
system of the form
\begin{equation}
\label{eq:matrixeq}
(\mathbf{\Tilde{H}} - ik \mathbf{L} - k^{2} \mathbf{O})
 \vec{c} = 0,
\end{equation}
\noindent
where \(L_{j,j'}\) is the surface matrix
\(y_{j}(R_{0}) y_{j'}(R_{0})\), \(\Tilde{H}_{j,j'}\) is the matrix
\(2\mu H_{j,j'}+y_{j}(R_{0}) \frac{d}{dr}y_{j'}(R_{0})\), and \(\mathbf{O}\)
is the overlap matrix for the spline basis set.

This equation is manifestly nonlinear, but the method of Tolstikhin
\emph{et al.}~allows it to be ``linearized'' by recasting it as a new
eigensystem in a basis of doubled dimension. (\cite{Friedman1956,Huestis1975}
discuss related techniques for solving differential equations where the
eigenvalue appears in a boundary condition.) We define \(d_{i}=ik c_{i}\),
yielding a trivial second equation \(ik\mathbf{O}\vec{c}=\mathbf{O}\vec{d}\).
Substituting this into the original eigenequation now gives a linear equation
in the doubled basis space
\begin{equation}
\label{eq:matrixeq2}
\begin{pmatrix} \mathbf{\Tilde{H}} & 0 \\ 0 & -\mathbf{O} \end{pmatrix}
 \begin{pmatrix} \vec{c} \\ \vec{d} \end{pmatrix} =
 ik \begin{pmatrix} \mathbf{L} & -\mathbf{O} \\ 
  -\mathbf{O} & 0 \end{pmatrix}
 \begin{pmatrix} \vec{c} \\ \vec{d} \end{pmatrix}.
\end{equation}
This is an equation for the eigenvalue \(\lambda=ik\), giving \(2N\) solutions
lying either on the (Re \(\lambda\))-axis or in conjugate pairs in the right
half of the complex plane.

In their work, Tolstikhin \emph{et al.}~used completeness properties of the
Siegert state set to construct a Green's function of the Hamiltonian, a
scattered solution, and the associated scattering matrix, for a variety of
single channel model problems. For systems with multiple channels, we instead
appeal to the well-understood machinery of multichannel quantum defect theory
(MQDT).

For resonance series corresponding to high electronically excited
intermediates (Rydberg states) of diatomic molecules, the most natural
description of the system is one with quantum defect parameters defined in
terms of a fixed internuclear distance \(R\) and a well-defined projection of
the orbital angular momentum \(\Lambda\) onto the axis of symmetry. This is
because the electron spends most of its time far from the nuclear core, and
when it does penetrate into the core, it gains enough speed from falling
through the Coulomb potential that the nuclei are essentially frozen on the
time scale of its motion. The quantum defect functions \(\mu_{\Lambda}(R)\) in
this representation, the so-called ``body-frame'', may either be calculated
from highly accurate \emph{ab initio} techniques, or extracted from a
semi-empirical fitting of experimental data \cite{Jungen1997}. In order to
connect them with the true asymptotic ionization channels defined in terms of
Siegert pseudostates of the residual core, \(j=\{v^{+},N^{+}\}\), a
\emph{frame transformation} must be performed \cite{Fano1970,Greene1985},
where \(N^{+}\) is the ionic rotational momentum, and \(v^{+}\) is the
vibrational quantum number of the pseudostates. In our procedure, we directly
evaluate the S-matrix by the frame transformation integral
\begin{equation}
\label{eq:ftran}
\begin{split}
S_{j,j'} =
 & \sum_{\Lambda} \langle N^{+}|\Lambda \rangle
 \int_{0}^{R_{0}} \phi_{j}(R) e^{2i\pi \mu_{\Lambda}(R)} \phi_{j'}(R) \, dR
 \, \langle \Lambda|N^{+'} \rangle \\
 & + i \sum_{\Lambda} \langle N^{+}|\Lambda \rangle
 \frac{\phi_{j}(R_{0}) e^{2i\pi \mu_{\Lambda}(R_{0})} \phi_{j'}(R_{0})}
 {k_{j}+k_{j'}}
 \, \langle \Lambda|N^{+'} \rangle. \\
\end{split}
\end{equation}
\noindent
The surface term in (6) also arises in the orthonormality relation
\cite{Tolstikhin1998}. A similar transformation converts the body-frame
transition dipole elements \(D_{\Lambda}(R)\) into reduced dipole matrix
elements in the same S-matrix representation,
\begin{equation}
\label{eq:ftran_dip}
\begin{split}
D^{S}_{j} = \,
 & (2J+1) \sum_{\Lambda} \langle \Lambda|J_{0} \rangle ^{(J)} 
 \, \langle \Lambda|N^{+} \rangle \\
 & \times \int_{0}^{R_{0}} \phi_{0}(R) D_{\Lambda}(R)
 e^{i\pi \mu_{\Lambda}(R)} \phi_{j}(R) \, dR. \\
\end{split}
\end{equation}
\noindent
Here \(\phi_{0}(R)\) is the initial vibrational wavefunction, and  \(J_{0}\)
and \(J\) are the total angular momenta of the initial and final states of the
system, respectively. Note that the Siegert pseudostates are \emph{never}
conjugated in these expressions, even when they formally belong to the dual
(``bra'') space. In particular, this means that the quantity labeled as
\(\vec{D}^{S \dagger}\) below is calculated by conjugating only
\(e^{i\pi \mu_{\Lambda}(R)}\) in the definition above, and not the dipole
matrix elements directly.

At this stage of the calculation no information about the long-range behavior
of the channels has yet been included, and since the body-frame quantum defects
are nearly energy independent, the resulting S-matrix is typically a smooth and
fairly weak function of energy. The method of \emph{channel elimination}
\cite{Seaton1983,Aymar1996} systematically eliminates flux in all electronic
channels below the energy threshold for electron escape (the ``closed-channel
subspace'') to form a ``physical'' S-matrix \( \mathbf{S}^{phys} \), by taking
the proper linear combination of short-range solutions that ensures
exponential decay at infinity. For a long-range Coulomb potential, this
procedure gives
\begin{equation}
\label{eq:channelim}
\mathbf{S}^{phys} = \mathbf{S}_{oo}^{}
 -\mathbf{S}_{oc}^{} (\mathbf{S}_{cc}^{} - e^{-2i\beta})^{-1}
 \mathbf{S}_{co}^{}.
\end{equation}
\noindent
Here, \(\beta\) is a diagonal matrix of the usual Coulomb long-range phase
parameter \(\pi (\nu_{j})\) where \(\nu_{j}\) is the (possibly complex)
effective quantum number in the \(j\)th channel, \(\mathbf{S}\) is the
scattering matrix, and the subscripts indicate partitions of the matrices into
closed and open subspaces \cite{Aymar1996}.

For a Siegert state basis, this physical scattering matrix is in general
\emph{not} unitary, but rather subunitary, reflecting the loss of flux at the
boundary \(R_{0}\) via coupling to the Siegert pseudo-continuum states. It may
be used to calculate partial cross-sections by means of conventional formulae,
but with the departure from unitarity,
\(1 - \sum_{j} |\mathbf{S}^{phys}_{j,j'}|^{2}\), identified as the probability
\(|\mathbf{S}^{phys}_{d,j'}|^{2}\) for scattering into the dissociative
continuum. This method also provides all quantities necessary to find the
partial photoionization cross-section into any open channel, \(\sigma_{j}\);
see Eq. 2.59 of \cite{Aymar1996} for further details. The contributions from
all open channels may then be summed to give the total cross-section for
photoionization.

Alternatively, the total photoabsorption cross-section may be found directly
from a \emph{``preconvolution'' formula} first derived by Robicheaux to handle
the energy smoothing of densely spaced resonances
\cite{Robicheaux1993,Granger2000},
\begin{equation}
\label{eq:preconv}
\begin{split}
\sigma_{total}(E) = 
 & \frac{4 \pi^{2} \alpha \omega}{3 (2J_{0}+1)}
 \text{Re} \, \vec{D}^{S \dagger} \\
 & \qquad \left[\mathbf{1} - \mathbf{S} e^{2i\beta}\right]^{-1}
 \left[\mathbf{1} + \mathbf{S} e^{2i\beta}\right] \vec{D}^{S} \\
\end{split}
\end{equation}
\noindent
where Re signifies taking the real part of everything that follows, and the 
\(\dagger\) here conjugates only the operator, not the entire matrix
element. The diagonal matrix written as \(e^{-2i\beta}\) has a nontrivial
definition in terms of the quantum defect parameters, it may be approximated
quite well by taking \(\beta_{j}= i \infty\) for ``closed'' channels with
\((E) < \text{Re}\,E_{j}\), and \(\beta_{i}=\pi \nu_{j}\) for ``open''
channels with \((E) > \text{Re}\,E_{j}\). Here \(E\) is the total energy of
the system, \(E_{j}\) is the threshold energy for channel \(j\), and
\(\nu_{j}=1/\sqrt{2(E_{j}-E)}\) on the branch where Im~\(\nu > 0\).
The utility of this expression lies in recognizing that the value of the
cross-section at a complex energy in the above formula is equivalent to the
cross-section at a real energy, smoothed over a channel-dependent width
\(\Gamma_{j} = 2 \, \text{Im}\,\epsilon_{j}\). Within the Siegert state
formulation, the electron energy \(\epsilon_{j} = E - E_{j}\) will naturally
take on a complex value in any channel where the channel eigenenergy \(E_{j}\)
is itself complex, while \(E\) remains real.

Given \(\mathbf{S}\) and \(\epsilon_{j}\), either of the two cross-section
formulae above may be evaluated, with appropriate allowances for the
possibility of complex energy eigenvalues. Note that the first procedure
simply gives a sum over the flux into specific ionization channels, while the
second gives a single value for the total photoabsorption cross-section.
This means that the latter will contain information about the solution
wavefunction along the \(R=R_{0}\) boundary not contained in any of the open
ionization channels. In general, the value of \(\sigma_{total}\) will be equal
to \emph{or greater than} the sum over the individual \(\sigma_{j}\), and any
difference may be attributed to the effect of coupling to high-lying Siegert
states in the continuum. Thus, the difference between these two formulae at
any energy provides the dissociative cross-section.

In order to test the validity of this hypothesis, we began by defining a set
of Siegert pseudostates for the \(H_{2}^{+}\) internuclear potential. The
eigensolutions fall into three classes, as shown in Fig.~1. Those lying
along the positive (Im~\(k\))-axis are associated with negative eigenenergies
on the physical sheet of the \(E\)-plane, the bound states of the potential.
These are the channel thresholds to which the Rydberg autoionization series of
the ionization spectrum converge, and so we include all of their states. The
solutions along the negative (Im~\(k\))-axis lie on the unphysical energy
sheet, and we reject them as antibound states arising from the doubling of
the dimension space. The remainder of the solutions fall above and below the
(Re~\(E\))-axis, corresponding to conjugate solution pairs of the eigenvalue
parameter \(\lambda=ik\). We select only those with negative Im~\(E_{j}\),
ensuring that they obey outgoing wave boundary conditions. (This amounts to
selecting only those states that contribute to the retarded, rather than the
advanced, Green's function. See \cite{Tanabe2001} for related discussion.) For
MQDT matrix elements it is also acceptable to reject states lying very high in
the continuum, since their Franck-Condon overlap with the bound states is
negligible.

Tolstikhin \emph{et al.}~discuss the unusual completeness relation obeyed by
the full set of Siegert pseudostates, with an additional factor of \(2\). Our
restricted subset of Siegert pseudostates does not, of course, obey that
doubled completeness relation. We have confirmed through numerical tests,
however, that our restricted subset behaves like a complete set, to at least
\(10^{-12}\) accuracy, for representing either \(L^{2}\) functions confined
within the boundary or functions with purely outgoing wave character at the
boundary. For an impressive demonstration (in a somewhat different context) of
the convergence properties of a similarly truncated Siegert basis also used to
describe smooth continuum physics, see \cite{Seideman1991}.

In the region of the ungerade H$_{2}$ spectrum between 127200 and 127800
cm\(^{-1}\) there are several strongly predissociated resonances, members of
the \(np\pi,v^{+}=8\) and \(np\pi,v^{+}=5\) series. In each case, our
calculated spectrum correctly reproduces them in the total absorption
cross-section, but shows them as weak or absent in the ionization. Comparisons
of our results with other theoretical and experimental values
\cite{GlassMaujean1987,Jungen1997} for the relative yields of selected
resonances appear in Table~I. Note particularly that we are able to correctly
describe the strong rotational dependence of the \(4p\pi,v^{+}=5\) branching
ratio, a nontrivial consequence of subtle channel interactions.

As a test of the method in an entirely different energy regime we considered
the problem of dissociative photoionization, a three-body breakup channel
accessible only at much higher energies. Experimental measures of the ratio
between pure ionization and dissociative ionization have been performed since
the 1970s by a number of researchers \cite{Browning1973,Backx1976,Chung1993},
along with at least one early theoretical calculation \cite{Ford1975}. Since
our ionization spectrum is a sum over individual channels, we can easily
distinguish between contributions from channels above and below the
dissociative threshold. Our results, plotted against those of past experiment
and theory, are presented in Fig.~2.

Finally, we have performed a model calculation demonstrating the utility of
our method for treating dissociative recombination, particularly in systems
where indirect channels (those involving scattering into intermediate
autodissociating Rydberg states) play an important role. Fig.~3 shows
the dissociative recombination spectrum of a simplified H$_{2}$ model
potential (neglecting rotation and with R-independent quantum defects),
compared with the familiar approximation of O'Malley for smooth background
scattering by direct processes \cite{OMalley1966}. Our spectrum accurately
reproduces this background, and also describes complex interference effects
from the series of resonances converging to each Rydberg threshold.

Some aspects of the Siegert MQDT method remain poorly understood, and would
benefit from greater clarification. For example, the utility of a subset of
the Siegert basis for MQDT depends on ability of that subset to represent
all energetically accessible regions of the continuum. While this requirement
appears from our numerical tests to be reasonably easy to satisfy, we have not
yet rigorously derived it from the relevant completeness relations. Also, it
is not presently clear how to extend the energy-smoothed formula to include
non-Coulombic long-range electronic potentials.

Other avenues of investigation could provide insight concerning the
applicability of our method to more complex systems. Polyatomic molecules, for
example, might be handled either by reduction to hyperspherical coordinates
\cite{Kokoouline2001}, or by a multidimensional generalization of the Sierget
state boundary conditions on an arbitrary hypersurface. Since our method
yields a full solution to the Schr\"odinger equation along the boundary at
R$_{0}$, it should also be possible to project onto the continuum functions of
different dissociative channels, and explicitly resolve partial dissociation
cross-sections \cite{Yaris1979}. Even in its current form, however, we believe
the Siegert MQDT method offers a simple description of the flux escaping into
dissociative channels, by working within a channel basis that obeys a
physically motivated boundary condition.

%Acknowledgements
This work is supported in part by a grant from the National Science Foundation.
We thank B. Esry for assistance in the early stages. Discussion with R. Santra,
M. Baertschy, M.~S. Child, C.~W. McCurdy, J.~M. Hutson, T.~N. Rescigno, and
B.~I. Schneider has also been helpful.

%\newpage
%\begin{references}

%\newpage
%FIGURES

\begin{minipage}{3.4in}
\noindent
TABLE I. Photoionization and photodissociation yields for select
\emph{ungerade} resonances in H\(_{2}\) for which the relative yields have been
experimentally observed \cite{GlassMaujean1987}.

\begin{tabular}{r|c|c|c|c}
\hline
\multicolumn{1}{c|}{State} & 
 \multicolumn{1}{c|}{Source} & 
 \multicolumn{1}{c|}{Energy} & 
 \multicolumn{1}{c|}{\% Ion.} & 
 \multicolumn{1}{c}{\% Diss.} \\ \hline
\(3p\pi,v^{+}=8,R(0)\) & Observed
 & 127248.2 & 10(5) & 95(5) \\
& Theory\(^{\mbox{\cite{Jungen1997}}}\) & 127246.9 & \mbox{ 1} & 99 \\
& Present & 127242.2 & \mbox{ 1} & 99 \\
\hline
\(5p\sigma,v^{+}=4,R(0)\) & Observed
 & 127599.4 & 90(10) & 10(10) \\
& Theory\(^{\mbox{\cite{Jungen1997}}}\) & 127602.2 & 88 & 12 \\
& Present & 127606.8 & 76 & 24 \\
\hline
\(4p\pi,v^{+}=5,R(0)\) & Observed
 & 127667.6 & 82(5) & 18(5) \\
& Theory\(^{\mbox{\cite{Jungen1997}}}\) & 127665.4 & 93 & \mbox{ 7} \\
& Present & 127666.6 & 97 & \mbox{ 3} \\
\hline
\(4p\pi,v^{+}=5,R(1)\) & Observed
 & 127599.4 & 30(10) & 70(10) \\
& Theory\(^{\mbox{\cite{Jungen1997}}}\) & 127758.4 & 17 & 83 \\
& Present & 127759.5 & 29 & 71 \\
\end{tabular}
\end{minipage}

%\newpage
\begin{figure}
\epsfxsize=3.4in \epsfbox{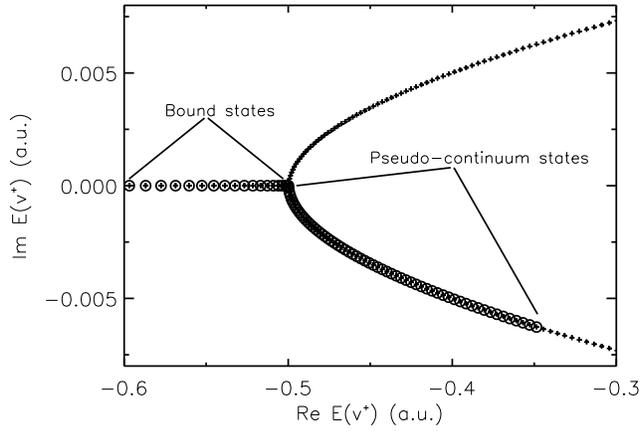}
\caption{Distribution of H$_{2}^{+}$ vibrational Siegert pseudostate energies
in the complex energy plane for angular momentum N$^{+}$=1. Only the circled
states are included as channels in the scattering matrix.}
\label{fig:sps}
\end{figure}

\begin{figure}
\epsfxsize=3.4in \epsfbox{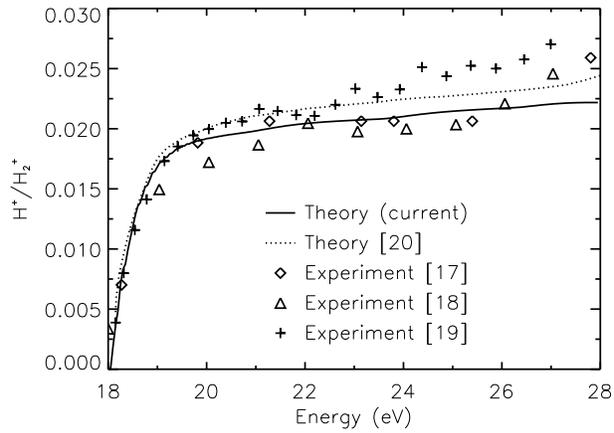}
\caption{Dissociative photoionization cross-section, as a ratio to the total
photoionization cross-section.}
\label{fig:dixsec}
\end{figure}

\begin{figure}
\epsfxsize=3.4in \epsfbox{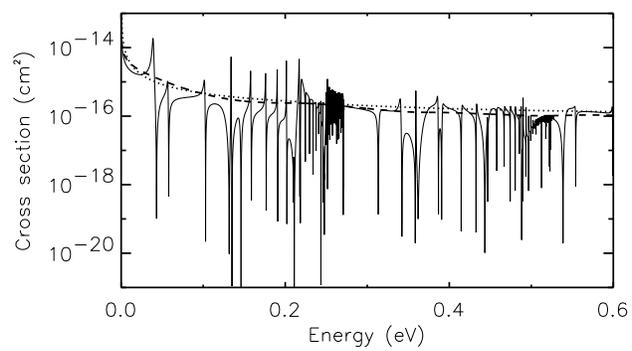}
\caption{Dissociative recombination cross-section for the model potential,
unconvolved (solid) and convolved with a Lorentzian of width 0.1 eV (dashed),
compared to that resulting from the O'Malley formula (dotted).}
\label{fig:drxsec}
\end{figure}

%\begin{figure}[tbp]
%\centerline{\epsfxsize=3.4in \epsfbox{Fig1.eps}}
%\centerline{\epsfxsize=3.4in \epsfbox{Fig2.eps}}
%\caption{

%\label{Fig1}
%\end{figure}

\end{document}